\documentclass[twocolumn]{aastex631}
\usepackage{graphicx}
\usepackage{float}
\usepackage{amsmath}
\usepackage{amssymb}
\usepackage{subfigure}
\usepackage{cases}
\usepackage{mathrsfs}
\usepackage[flushleft]{threeparttable}
\usepackage{amssymb}
\usepackage{pifont}
\usepackage{epstopdf}
\usepackage{xcolor}
\usepackage[all]{hypcap}
\usepackage{mwe,tikz}
\usepackage{bm}
\usepackage{multirow}
\usepackage{longtable}
\usepackage{xspace}
\usepackage{rotating}
\usepackage{CJK}
\usepackage{url}
\usepackage{upgreek}
\usepackage{booktabs}
\usepackage{textcomp}
\usepackage{makecell}
\usepackage{enumitem}
\usepackage{csquotes}
\usepackage{soul}
\usepackage{tabularx}
\usepackage{threeparttable}

\newcommand{\e}{\text{e}}

\newcommand{\p}{\partial}

\begin{document}

 % \title{Bridging the Gap: GRB 230812B—A Three-Second Supernova-Associated Burst Detected by The GRID CubeSat Mission}
\title{Bridging the Gap: GRB 230812B --- A Three-Second Supernova-Associated Burst Detected by the GRID Mission}

% Tier 1 :

\correspondingauthor{Bin-Bin Zhang, Hua Feng, Ming Zeng}
\email{bbzhang@nju.edu.cn; hfeng@ihep.ac.cn; zengming@tsinghua.edu.cn}

\author[0009-0007-9068-2752]{Chen-Yu Wang}
\affiliation{Department of Astronomy, Tsinghua University, Beĳing 100084, China}

\author[0000-0002-5596-5059]{Yi-Han Iris Yin}
\affiliation{School of Astronomy and Space Science, Nanjing University, Nanjing 210093, China}
\affiliation{Key Laboratory of Modern Astronomy and Astrophysics (Nanjing University), Ministry of Education, China}

\author[0000-0003-4111-5958]{Bin-Bin Zhang}
\affiliation{School of Astronomy and Space Science, Nanjing University, Nanjing 210093, China}
\affiliation{Key Laboratory of Modern Astronomy and Astrophysics (Nanjing University), Ministry of Education, China}
\affiliation{Purple Mountain Observatory, Chinese Academy of Sciences, Nanjing, 210023, China}

\author[0000-0001-7584-6236]{Hua Feng}
\affiliation{Key Laboratory of Particle Astrophysics, Institute of High Energy Physics, Chinese Academy of Sciences, 100049, Beijing, China}

\author[0000-0001-7584-6236]{Ming Zeng}
\affiliation{Key Laboratory of Particle Astrophysics, Institute of High Energy Physics, Chinese Academy of Sciences, 100049, Beijing, China}
\affiliation{Department of Engineering Physics, Tsinghua University, Beijing, 100084, People’s Republic of China}

\author{Shao-Lin Xiong}
\affiliation{Key Laboratory of Particle Astrophysics, Institute of High Energy Physics, Chinese Academy of Sciences, 100049, Beijing, China}

% Tier 2 :

\author[0000-0002-7439-6621]{Xiao-Fan Pan}
\affiliation{Department of Engineering Physics, Tsinghua University, Beijing, 100084, People’s Republic of China}

\author[0000-0002-5485-5042]{Jun Yang}
\affiliation{School of Astronomy and Space Science, Nanjing University, Nanjing 210093, China}
\affiliation{Key Laboratory of Modern Astronomy and Astrophysics (Nanjing University), Ministry of Education, China}

\author{Yan-Qiu Zhang}
\affiliation{Key Laboratory of Particle Astrophysics, Institute of High Energy Physics, Chinese Academy of Sciences, 100049, Beijing, China}
\affiliation{University of Chinese Academy of Sciences, Chinese Academy of Sciences, Beijing 100049, China}

\author{Chen Li}
\affiliation{Department of Engineering Physics, Tsinghua University, Beijing, 100084, People’s Republic of China}

\author[0009-0008-2841-3065]{Zhen-Yu Yan}
\affiliation{School of Astronomy and Space Science, Nanjing University, Nanjing 210093, China}
\affiliation{Key Laboratory of Modern Astronomy and Astrophysics (Nanjing University), Ministry of Education, China}

\author{Chen-Wei Wang}
\affiliation{Key Laboratory of Particle Astrophysics, Institute of High Energy Physics, Chinese Academy of Sciences, 100049, Beijing, China}
\affiliation{University of Chinese Academy of Sciences, Chinese Academy of Sciences, Beijing 100049, China}

\author{Xu-Tao Zheng}
\affiliation{Department of Engineering Physics, Tsinghua University, Beijing, 100084, People’s Republic of China}

\author{Jia-Cong Liu}
\affiliation{Key Laboratory of Particle Astrophysics, Institute of High Energy Physics, Chinese Academy of Sciences, 100049, Beijing, China}
\affiliation{University of Chinese Academy of Sciences, Chinese Academy of Sciences, Beijing 100049, China}

\author{Qi-Dong Wang}
\affiliation{Department of Engineering Physics, Tsinghua University, Beijing, 100084, People’s Republic of China}

\author{Zi-Rui Yang}
\affiliation{Department of Engineering Physics, Tsinghua University, Beijing, 100084, People’s Republic of China}

\author{Long-Hao Li}
\affiliation{Department of Engineering Physics, Tsinghua University, Beijing, 100084, People’s Republic of China}

\author{Qi-Ze Liu}
\affiliation{Department of Engineering Physics, Tsinghua University, Beijing, 100084, People’s Republic of China}

\author{Zheng-Yang Zhao}
\affiliation{Weiyang College, Tsinghua University, Beijing, 100084, People’s Republic of China}

% Tier 3 :
\author{Bo Hu}
\affiliation{Weiyang College, Tsinghua University, Beijing, 100084, People’s Republic of China}

\author{Yi-Qi Liu}
\affiliation{Weiyang College, Tsinghua University, Beijing, 100084, People’s Republic of China}

\author{Si-Yuan Lu}
\affiliation{Weiyang College, Tsinghua University, Beijing, 100084, People’s Republic of China}

\author{Zi-You Luo}
\affiliation{School of Aerospace Engineering, Tsinghua University, Beijing, 100084, People’s Republic of China}

\author{Ji-Rong Cang}
\affiliation{Star Detect CO., LTD., Beijing 100190, People's Republic of China}

\author{De-Zhi Cao}
\affiliation{Star Detect CO., LTD., Beijing 100190, People's Republic of China}

\author{Wen-Tao Han}
\affiliation{Department of Computer Science and Technology, Tsinghua University, Beijing, 100084, People’s Republic of China}

\author{Li-Ping Jia}
\affiliation{Star Detect CO., LTD., Beijing 100190, People's Republic of China}

\author{Xing-Yu Pan}
\affiliation{Star Detect CO., LTD., Beijing 100190, People's Republic of China}

\author{Yang Tian}
\affiliation{Key Laboratory of Particle Astrophysics, Institute of High Energy Physics, Chinese Academy of Sciences, 100049, Beijing, China}
\affiliation{Department of Engineering Physics, Tsinghua University, Beijing, 100084, People’s Republic of China}

\author{Ben-Da Xu}
\affiliation{Department of Engineering Physics, Tsinghua University, Beijing, 100084, People’s Republic of China}

\author{Xiao Yang}
\affiliation{Weiyang College, Tsinghua University, Beijing, 100084, People’s Republic of China}

\author{Zhi Zeng}
\affiliation{Key Laboratory of Particle Astrophysics, Institute of High Energy Physics, Chinese Academy of Sciences, 100049, Beijing, China}
\affiliation{Department of Engineering Physics, Tsinghua University, Beijing, 100084, People’s Republic of China}

\author{GRID Collaboration}

\begin{abstract} GRB 230812B, detected by the Gamma-Ray Integrated Detectors (GRID) constellation mission, is an exceptionally bright gamma-ray burst (GRB) with a duration of only 3 seconds. Sitting near the traditional boundary ($\sim$ 2 s) between long and short GRBs, GRB 230812B is notably associated with a supernova (SN), indicating a massive star progenitor. This makes it a rare example of a short-duration GRB resulting from stellar collapse. Our analysis, using a time-evolving synchrotron model, \textcolor{black}{suggests that the burst has an emission radius of approximately $10^{14.5}$~cm}. We propose that the short duration of GRB 230812B is due to the combined effects of the central engine's activity time and the time required for the jet to break through the stellar envelope. Our findings provide another case that challenges the conventional view that short-duration GRBs originate exclusively from compact object mergers, demonstrating that a broader range of durations exists for GRBs arising from the collapse of massive stars. \end{abstract}

\keywords{Gamma-ray bursts; Radiation mechanism}

\section{Introduction}

Despite the traditional classification of gamma-ray bursts (GRBs) into short-duration (Type I) and long-duration (Type II) categories based on their durations and progenitors \citep{Kouveliotou1993, Woosley2006}, the boundary between these two types remains an open question. Recent observations of GRBs with atypical durations—such as minute-duration Type I GRBs \textcolor{black}{\citep[e.g., GRB 060614, GRB 211211A, GRB 230307A, ][]{Gehrels2006, Yang2022, Sun2023}} and second-duration Type II GRBs \citep[e.g., GRB 090426, GRB 200826A][]{Antonelli2018, Zhang2021}—have further blurred this distinction \citep{Zhang2021}. In the latter case, the short observed durations are often interpreted as resulting from the ``tip-of-the-iceberg'' effect, a baryon-loaded matter-dominated jet, or the time required for the jet to break through the stellar envelope. If such interpretation holds, it suggests a broader range of short durations for Type II GRBs, beyond the typical 2-second threshold \citep{Bromberg2012, Lu2014}. Although such peculiar GRBs are limited in number, they are essential for understanding the central engine timescales and the physics governing GRB durations. In this context, we report GRB 230812B, a Type II, SN-associated GRB with a duration of approximately 3 seconds, which further bridges the gap between the traditional categories and provides new insights into GRB physics.

GRB 230812B was detected by our Gamma-Ray Integrated Detectors (GRID\footnote{GRID has launched nine payloads to date, with seven currently operational. Covering over 80\% of the celestial sphere in real time, GRID is particularly adept at detecting transient events such as GRBs. To date, it has confirmed over forty GRBs \citep{Wen2019, Wang2021, Zhang2023c, Wang2023}.}) constellation project, a student-led and cost-effective mission designed to monitor GRBs in the 10 keV to 2 MeV energy range. Notably, while other detectors such as \textit{Fermi}/GBM suffered from saturation due to the extreme brightness of GRB 230812B, GRID provided reliable data for comprehensive analysis.

In this paper, we present a detailed analysis of GRB 230812B, using dedicated data obtained from the GRID mission. Section~\ref{tem analysis} outlines the observations, data reduction, and temporal analysis. In Section~\ref{sep analysis}, we perform time-integrated and time-resolved spectral analyses, applying both statistical and physical models, including synchrotron emission modeling, to investigate the burst’s physical characteristics and emission mechanisms. To further contextualize GRB 230812B, we compare its properties with other GRBs in Section~\ref{placement}, particularly highlighting its similarities to GRB 200826A, a similarly short-duration, supernova-associated GRB. Finally, Section~\ref{explanation} provides a discussion on the possible explanations for GRB 230812B’s short duration, focusing on the role of jet dynamics and the central engine’s activity. We conclude with a summary of the key findings in Section~\ref{summary}. 

\section{Observations, Data Reduction and Analysis}

At 18:58:12 Universal Time on August 12, 2023, GRB 230812B was observed by multiple missions, including \textit{Fermi} \citep{Roberts2023a}, GECAM-C \citep{Xiong2023}, and GRID-05B—the fifth detector launched in 2020 as part of the GRID constellation \citep{Wen2019}. Subsequent follow-up observations revealed the burst's afterglow counterpart and confirmed the presence of an associated supernova at a redshift of 0.36 \citep{Postigo2023}.

\begin{figure}
 \label{fig: loc}
 \centering
 \includegraphics[width = 0.45\textwidth]{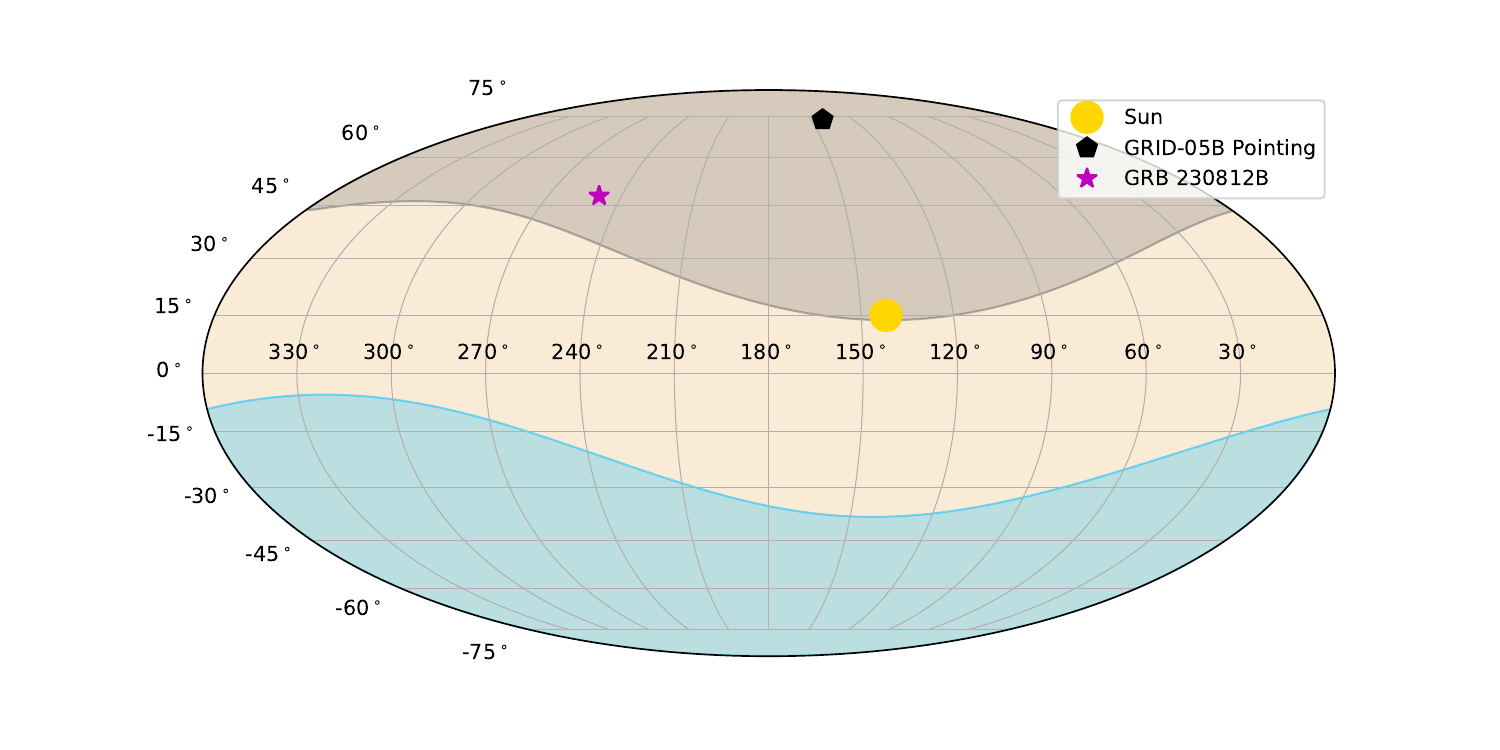}
 \caption{Skymap showing the location of GRB 230812B \textcolor{black}{(magenta star)} and the pointing direction of the GRID-05B detector \textcolor{black}{(black diamond)}. The gray-shaded region indicates areas \textcolor{black}{less than 60 degrees separated from GRID-05B's pointing}. The Sun is depicted as a gold \textcolor{black}{circle}, and the Earth's projection is shaded in blue. The large angular distance between both the GRB and GRID-05B's pointing direction relative to Earth further minimizes atmospheric reflection effects.}
%The GRB's position is marked by a magenta star, while the black diamond represents the pointing direction of GRID-05B.}
\end{figure}

Due to the exceptional brightness of GRB 230812B, \textit{Fermi}'s instruments experienced saturation and data loss during the observation period, rendering portions of its data unreliable \citep{Roberts2023b, Roberts2024}. Meanwhile, GECAM-C's measurements of this event were compromised by atmospheric reflection, leading to reduced spectral data quality\footnote{Per private communication with the GECAM Collaboration.}. In contrast, GRID-05B's data were unaffected by these issues, \textcolor{black}{owing to its observational conditions.} The \textcolor{black}{orbital location} of GRID-05B, as shown in Figure \ref{fig: loc}, ensured the reliability of its observations, given its significant but acceptable angular separation (approximately 49 degrees) from the burst's location, as well as the fact that Earth was outside its field of view. \textcolor{black}{{Additionally, the small effective area of the GRID detector allowed for effective photon collection without saturation, even from such a bright GRB.}} Consequently, while we referred to light curve data from \textcolor{black}{Fermi/GBM} and GECAM-C, the primary analysis of the temporal and spectral features of the burst was conducted using data from GRID-05B.

The GRID-05B data for GRB 230812B are publicly available at the National Space Science Data Center of the Chinese Academy of Sciences (\url{https://grid.nssdc.ac.cn/}). We downloaded the event data and generated the corresponding response matrix files using the most up-to-date calibration database, calculated by Geant 4 \citep{Wen2019}. We identified the burst using the method described by \citet{Zou2021} and performed the temporal and spectral analyses using our customized pipeline, which is in accordance with the methodologies described by \citet{Wang2021} and \citet{2023RAA....23k5023Z}.  Detailed results of these analyses are provided below.

\subsection{The Temporal Analysis}\label{tem analysis}

\begin{figure}
 \centering
 \includegraphics[width = 0.45\textwidth]{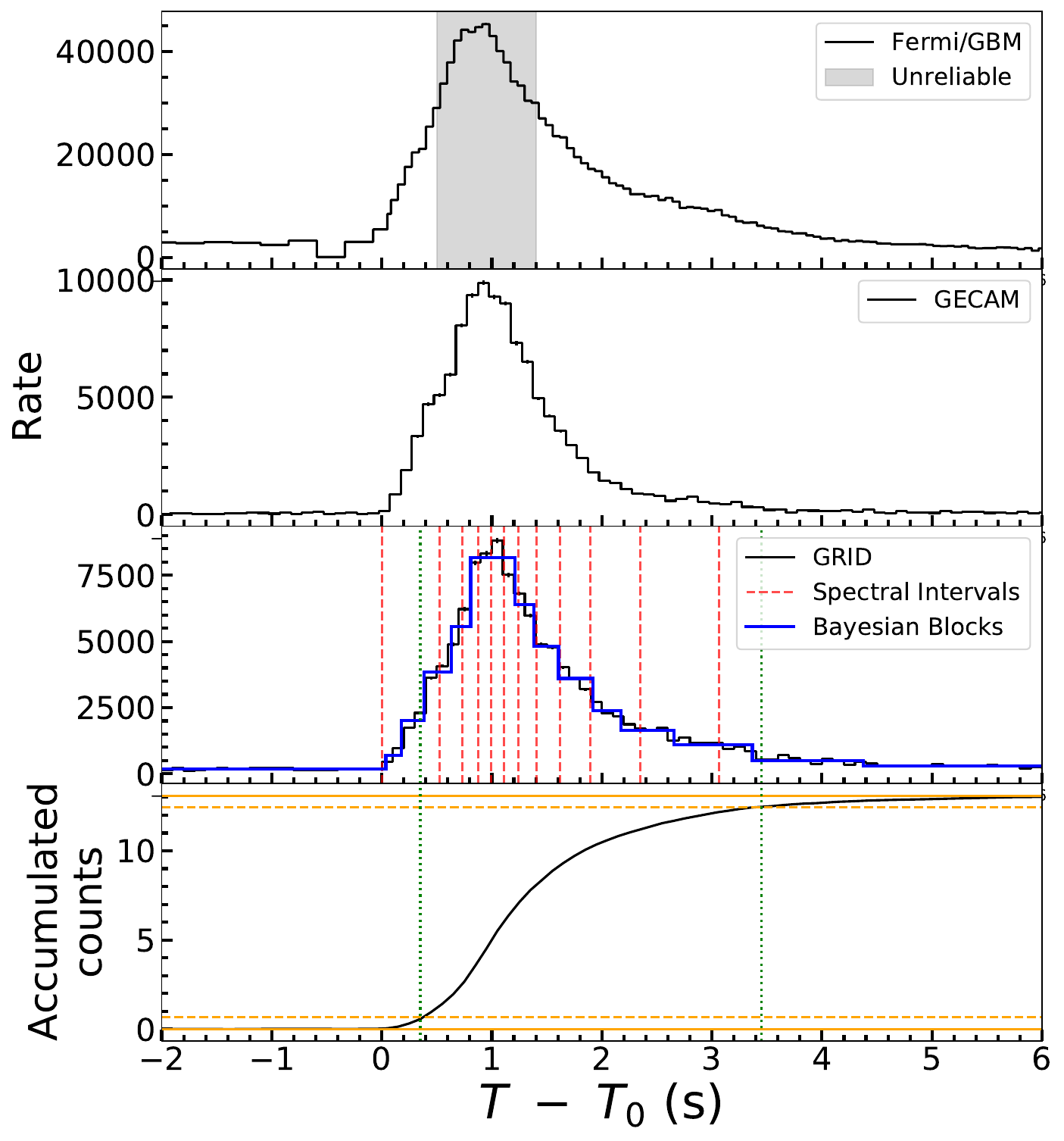}
\caption{Light curves and $T_{\rm{90}}$ calculation for GRB 230812B, shown with a bin size of 0.1 seconds. The top \textcolor{black}{three panels} display the light curves from three instruments: Fermi/GBM, GECAM, and GRID-05B, within the 10 to 1000 keV range. Due to its smaller effective area, the count rate from GRID-05B is significantly lower than those from the other two instruments, which avoids data saturation for this exceptionally bright GRB. The \textcolor{black}{third} panel shows the Bayesian Block partitions (blue line) and the time intervals selected for spectral analysis (red dashed lines). The bottom panel illustrates the accumulated counts curve (black line), with solid and dashed orange lines indicating the 0\% (5\%) and 100\% (95\%) levels, respectively. Vertical green dashed lines represent the $T_{\rm{90}}$ duration of the burst across the panels.}
 \label{fig: lccom}
\end{figure}

We present the light curves of GRB 230812B from the three instruments, GRID-05B, \textcolor{black}{Fermi/GBM}, and GECAM-C, in Figure \ref{fig: lccom}. These light curves were produced by processing the event data, which were downloaded from their public repositories, binned at 0.1 seconds, and corrected for barycentric motion. The consistency observed across all three instruments validates the data quality from each instrument involved in this observation.

\begin{figure}[ht!]
 \label{fig: t90_hist}
 \centering
 \includegraphics[width = 0.45\textwidth]{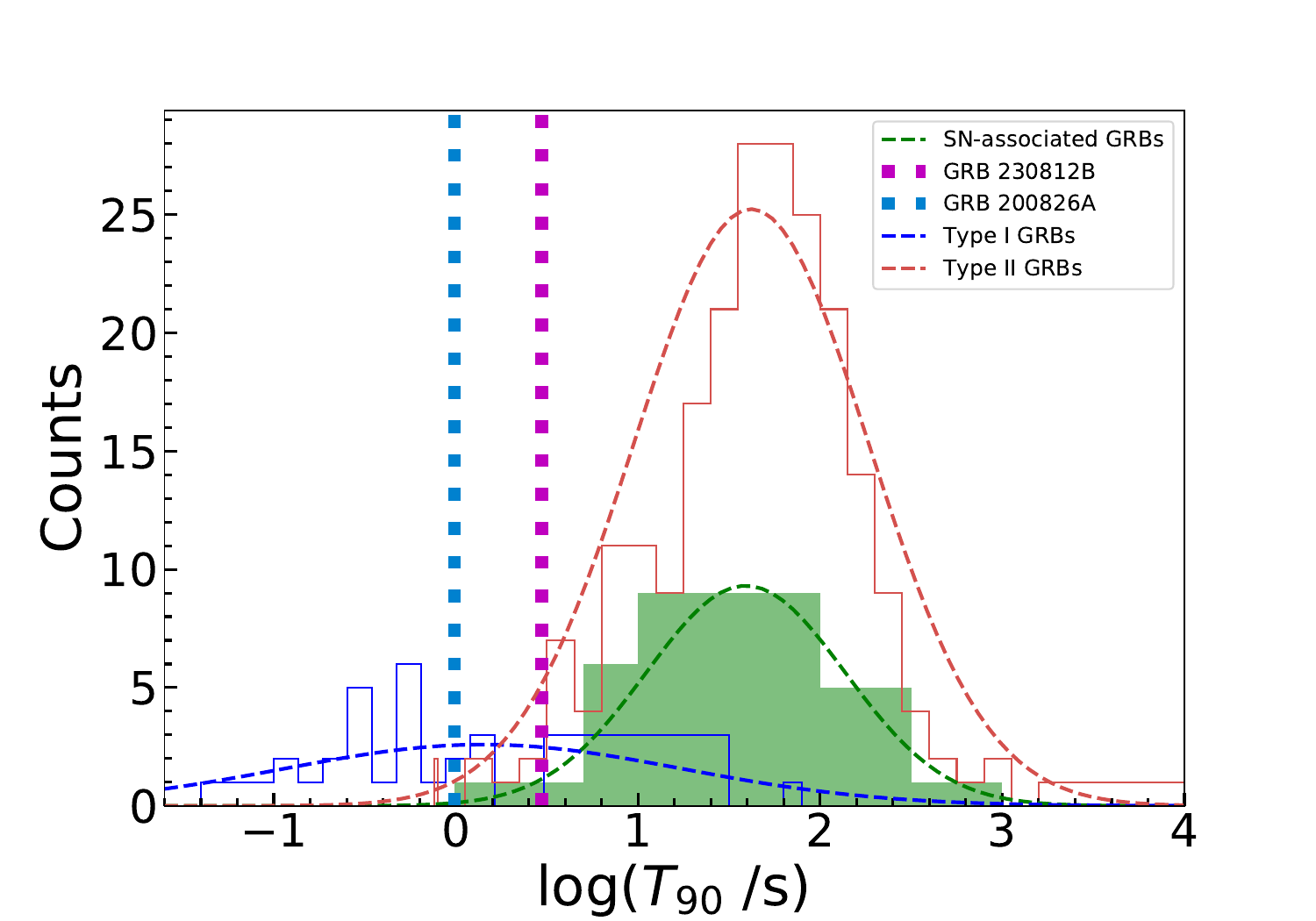}
\caption{$T_{\rm{90}}$  distributions for Type I GRBs, Type II GRBs, and GRB-SNe \citep{Frontera2024, Cano2017}. The blue, red, and green histograms correspond to the $T_{\rm{90}}$ distributions for Type I GRBs, Type II GRBs, and SN-associated GRBs, respectively, while the dashed lines represent the corresponding best-fit Gaussian distributions. The blue and magenta dotted lines indicate the $T_{\rm{90}}$ values for GRB 200826A and GRB 230812B.}

\end{figure}

Using the methods described by \citet{Yang2020a} and \citet{Yang2020b}, we calculated the $T_{\rm{90}}$ duration for GRB 230812B within the 10 keV–1000 keV energy band. This calculation was performed using event data from GRID-05B, with a bin size of 0.1 s, as illustrated in Figure \ref{fig: lccom} (middle panel). The resulting $T_{\rm{90}}$, as shown in Table \ref{tab: basicinfo}, is $2.97^{+0.03}_{-0.02}$ seconds, positioning GRB 230812B near the boundary between typical Type I and Type II GRBs, as illustrated in Figure \ref{fig: t90_hist}. Additionally, we plotted the $T_{\rm{90}}$ distribution of all supernova-associated GRBs (represented by the solid green-filled histogram in Figure \ref{fig: t90_hist}), revealing that GRB 230812B is the second shortest burst in this subsample, surpassed only by \textcolor{black}{GRB 200826A which has a 1-second duration.}

\begin{figure}
 \centering
 \includegraphics[width = 0.4\textwidth]{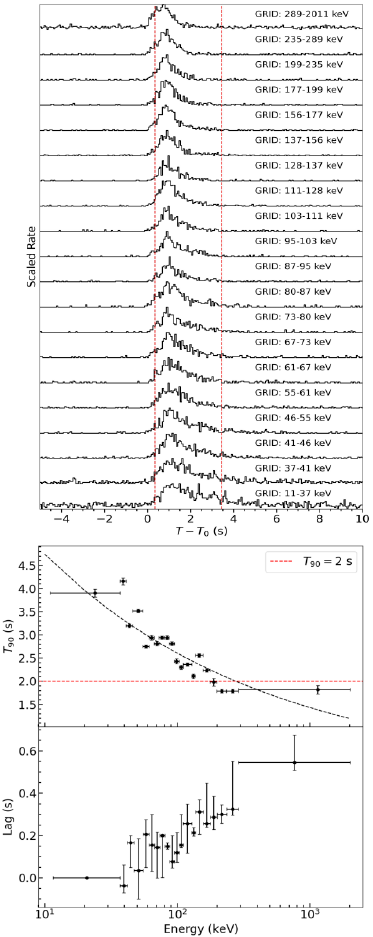}
\caption{Energy-dependent multi-wavelength light curves, $T_{\rm{90}}$ duration, and spectral lag calculations. The top panel shows the multi-wavelength light curves derived from GRID-05B across various energy bands. The middle panel presents the $T_{\rm{90}}$ values as a function of the midpoint energy for each of the energy-dependent light curves shown in the top panel. The red dashed line indicates the typical $T_{\rm{90}}$ division of 2 seconds. The bottom panel shows the energy-dependent spectral lags, with positive values indicating a delay in higher-energy photon arrival times relative to lower-energy photons. All error bars represent 1$\sigma$ uncertainties.}

 \label{fig: lag}
\end{figure}

By extracting light curves across different energy bands from GRID-05B data, we generated the multi-wavelength light curves shown in Figure \ref{fig: lag}. The pulse shapes exhibit a broadening trend toward lower energy bands, as reflected in the energy-dependent $T_{90}$ values in the middle panel—a characteristic feature typically associated with long GRBs. Moreover, we calculated the spectral lags between the lowest energy band and higher energy bands using the methodology outlined by \citet{Zhang2021}. The results, presented in the bottom panel of Figure \ref{fig: lag}, depict the lag relationship across the energy-dependent light curves. The significant spectral lags observed strongly suggest that GRB 230812B aligns with the characteristics of a Type II GRB.

\begin{table}[ht!]
 \centering
 \caption{Observational properties of GRB 230812B}
 \label{tab: basicinfo}
 \begin{tabular}{ll}
 \toprule
 \toprule
 Observed Properties & GRB 230812B\\
 \midrule
 $T_{\rm{90}}$ (s) & $2.97^{+0.03}_{-0.02}$\\
 Total duration (s) & $\sim$ 4.95\\
 $\alpha$ at peak & $-0.34^{+0.06}_{-0.15}$\\
 $E_{\rm{p}}$ at peak & $166.35^{+15.24}_{-12.68}$\\
 Time integrated $\alpha$ & $-0.95^{+0.08}_{-0.12}$\\
 Time integrated $E_{\rm{p}}$ (keV)& $346.91^{+8.70}_{-13.90}$\\
 Total fluence ($10^{-4}\ \rm{erg\ cm^{-2}}$)& $1.41^{+0.08}_{-0.13}$\\
 Peak flux ($10^{-4}\ \rm{erg\ cm^{-2}\ s^{-1}}$) & $1.90^{+0.05}_{-0.08}$\\
 Isotropic Energy ($10^{52}\ \rm{erg}$) & $9.76^{+0.11}_{-0.54}$\\
 Redshift & 0.36\\
 \bottomrule
 \end{tabular}
\end{table}

\subsection{The Spectral Analysis }\label{sep analysis}

Both time-integrated and time-resolved spectral analyses were performed. For each analysis, we first extracted the total spectrum for the given time slice and calculated the background spectrum by interpolating a 2nd- to 3rd-order polynomial from the source-free time intervals to the source region, as outlined in \citet{Zhang2018} and \citet{Yang2020b}. The response matrix file was generated following the methodology described in \textcolor{black}{\citet{Wen2019}}.

The time-integrated spectral analysis was performed over the entire burst period, from 0.01 to 4.97 seconds. We fitted the spectra using a cutoff power-law (CPL) model, formulated as $N(E)=K E^{\alpha} \exp\left(-\frac{E}{E_{\rm{c}}}\right)$, where $K$ is the normalization factor, $\alpha$ is the spectral index, and $E_c$ is the cutoff energy. As presented in Table \ref{tab: cpl}, the time-integrated spectrum is well-fitted with the CPL model, yielding $\alpha = -0.95^{+0.08}_{-0.12}$ and a peak energy of $E_{\rm{p}} = E_{\rm{c}}/(2 + \alpha) = 346.91^{+8.70}_{-13.90}$ keV. Based on this fit, the total fluence, calculated over the energy range from 10 to 1000 keV, is $8.79^{+0.08}_{-0.13} \times 10^{-4}\ \rm{erg\ cm^{-2}}$. Furthermore, with the measured redshift of 0.36 \citep{Postigo2023}, the isotropic energy release was computed to be $6.10^{+0.11}_{-0.54} \times 10^{53}\ \rm{erg}$.

\begin{table*}
 
 \tiny
 \centering
 \caption{The best-fit parameters of spectral fitting using CPL model}
 \label{tab: cpl}
 \resizebox{0.9\textwidth}{!}{%
 \begin{threeparttable}
 \begin{tabular}{cccccccc}
     \toprule
     \toprule
     ID & $t_{\rm{start}\ (s)}$ & $t_{\rm{end}\ (s)}$ & $\alpha$ & $\log (E_{\rm{p}}\ /\ \rm{keV})$ & $\log A_{\rm{GRID}}$\tnote{a} & $\frac{\rm{pgstat}}{\rm{dof}}$ & BIC\\
     \midrule
     1 & 0.01 & 0.48 & $-0.68^{+0.10}_{-0.15}$ & $3.06^{+0.21}_{-0.08}$ & $1.94^{+0.27}_{-0.2}$ & 72.51/61 & 84.98 \\
     2 & 0.48 & 0.69& $-0.46^{+0.10}_{-0.13}$ & $2.74^{+0.07}_{-0.04}$ & $2.01^{+0.24}_{-0.20}$ & 68.95/61 & 81.42 \\
     3 & 0.69 & 0.83& $-0.34^{+0.15}_{-0.11}$ & $2.63^{+0.04}_{-0.05}$ & $2.01^{+0.20}_{-0.27}$ & 66.11/61 & 78.59 \\
     4 & 0.83 & 0.95& $-0.25^{+0.14}_{-0.14}$ & $2.51^{+0.04}_{-0.03}$ & $2.03^{+0.23}_{-0.24}$ & 58.97/61 & 71.44 \\
     5 & 0.95 & 1.06& $-0.25^{+0.15}_{-0.15}$ & $2.45^{+0.04}_{-0.03}$ & $2.10^{+0.25}_{-0.26}$ & 65.40/61 & 77.88 \\
     6 & 1.06 & 1.19& $-0.41^{+0.14}_{-0.16}$ & $2.41^{+0.04}_{-0.03}$ & $2.36^{+0.26}_{-0.23}$ & 49.22/61 & 61.69 \\
     7 & 1.19 & 1.36& $-0.70^{+0.12}_{-0.15}$ & $2.46^{+0.05}_{-0.05}$ & $2.74^{+0.26}_{-0.20}$ & 56.32/61 & 68.80 \\
     8 & 1.36 & 1.57& $-0.64^{+0.14}_{-0.19}$ & $2.29^{+0.05}_{-0.03}$ & $2.64^{+0.29}_{-0.23}$ & 51.96/61 & 64.44 \\
     9 & 1.57 & 1.85& $-0.84^{+0.17}_{-0.17}$ & $2.23^{+0.04}_{-0.03}$ & $2.93^{+0.25}_{-0.27}$ & 61.88/61 & 74.36 \\
     10 & 1.85 & 2.31& $-0.85^{+0.19}_{-0.21}$ & $2.09^{+0.04}_{-0.03}$ & $2.82^{+0.31}_{-0.30}$ & 79.05/61 & 91.53 \\
     11 & 2.31 & 3.03& $-1.11^{+0.20}_{-0.29}$ & $1.94^{+0.03}_{-0.04}$ & $3.10^{+0.42}_{-0.28}$ & 69.58/61 & 82.06 \\
     12 & 3.03 & 4.97& $-1.72^{+0.32}_{-0.12}$ & $1.58^{+0.17}_{-0.15}$ & $3.55^{+0.17}_{0.45}$ & 85.31/61 & 97.79 \\
     \hline
     Total & 0.01 & 4.97 & $-0.95^{+0.08}_{-0.12}$ & $2.55^{+0.01}_{-0.02}$ & $2.61^{+0.21}_{-0.32}$ & 62.31/61 & 67.42 \\
     \bottomrule
 \end{tabular}
  \begin{tablenotes}
    \footnotesize
    \centering
    %\begin{minipage}{0.8\textwidth}
    \item[a] Normalization factor $A$ is in units of $\rm{ph\ keV^{-1}\ cm^{-2}\ s^{-1}}$.
    %\end{minipage}
  \end{tablenotes}
 \end{threeparttable}}
\end{table*}

The time-resolved spectral analysis was performed by dividing the burst into twelve time-dependent slices, as shown in Figure \ref{fig: lccom} and detailed in Table \ref{tab: cpl}. Each slice was selected to ensure that each of the 92 energy bands of GRID-05B contained at least ten photons, allowing for statistically robust spectral fitting. The fitting process for each slice utilized the Monte-Carlo fitting tool, McSpecFit \citep{Zhang2016, Zhang2018}, which facilitates accurate modeling of the spectral data. Five empirical models were applied in the analysis: the Band function \citep{Band1997}, blackbody (BB), multi-color blackbody (mBB), single power law (PL), and cutoff power law (CPL).

The goodness-of-fit for each model was evaluated using the Poisson data with a Gaussian background likelihood, normalized by the degrees of freedom (PGSTAT/dof, \citealt{Atwood2009}) and the Bayesian information criterion (BIC, \citealt{Schwarz1978}). Among the models considered, the CPL model consistently resulted in the smallest BIC values, signifying its preference as the best fit for the time-resolved spectra. The best-fit parameters obtained from the CPL model are presented in Table \ref{tab: cpl}.

\begin{figure}
 \centering
 \includegraphics[width = 0.45\textwidth]{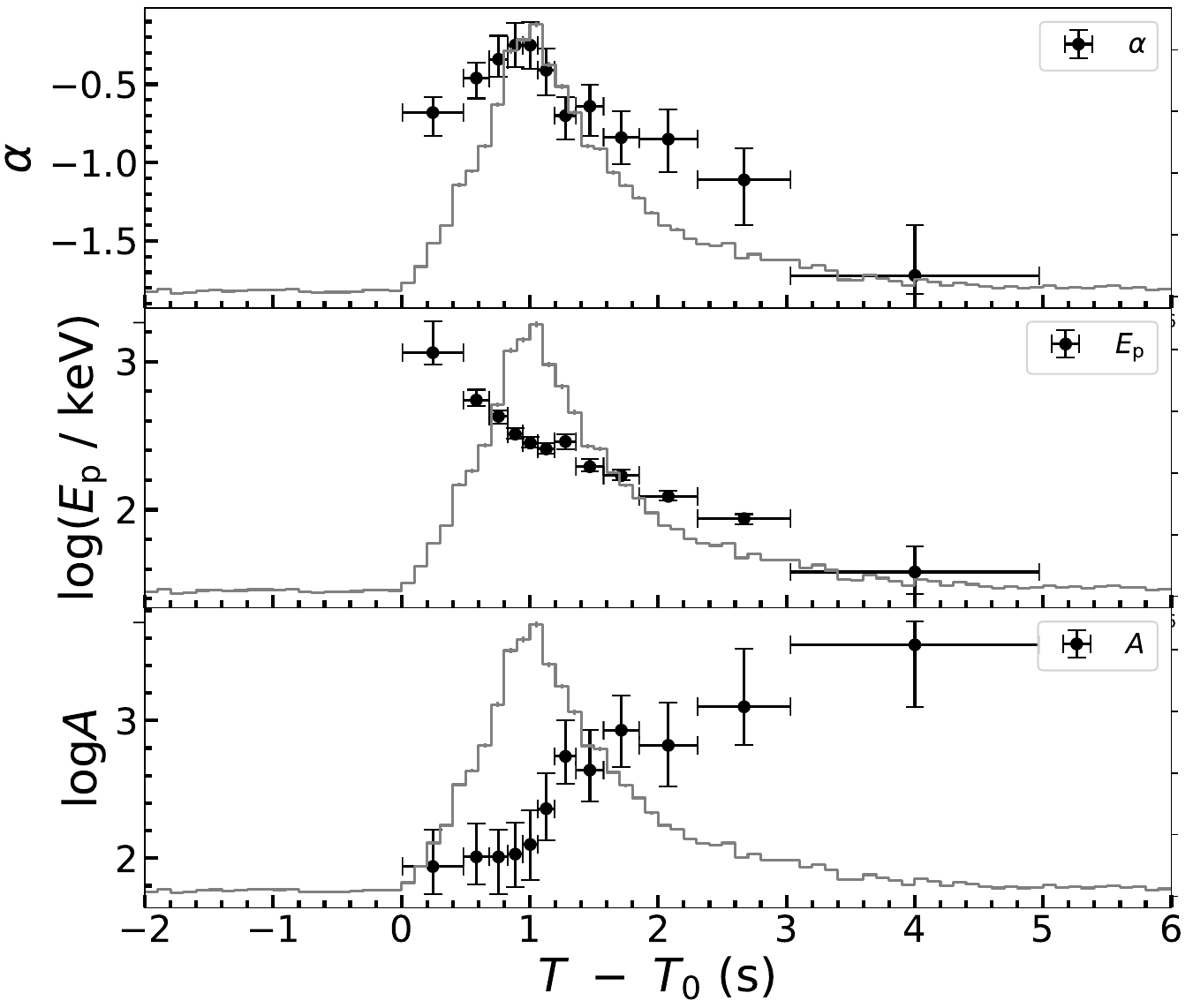}
 \caption{The spectral evolution of GRB 230812B represented by the best-fit parameters from the CPL model fits as a function of time. The top, middle, and bottom panels show the evolution of the low-energy photon index ($\alpha$), the peak energy ($E_{\rm{p}}$), and the normalization ($A$), respectively. The horizontal error bars represent the time intervals, and the vertical bars indicate the 1-$\sigma$ uncertainties of the best-fit parameters. In each panel, the light curve of the burst is also overplotted in the background for a \textcolor{black}{visual} comparison between the spectral evolution and the burst's temporal behavior.}

 \label{fig: cpl_param}
\end{figure}

As shown in Figure \ref{fig: cpl_param}, the spectral evolution of GRB 230812B follows a hard-to-soft trend in the peak energy $E_{\rm{p}}$. Simultaneously, the best-fit low-energy photon index $\alpha$ varies in correspondence with the flux level, indicating a coherent spectral behavior throughout the burst.

\begin{figure*}
 \centering
 \includegraphics[width = 0.9\textwidth]{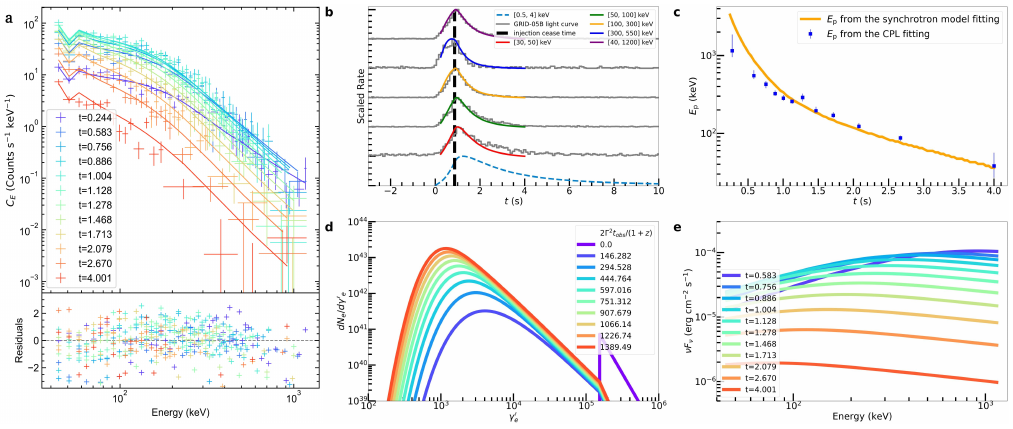}
 \caption{The fitting results of the synchrotron model for GRB 230812B. Panel a: Observed photon count spectra (data points with error bars) are plotted alongside the best-fit synchrotron model spectra (solid curves) \textcolor{red}{in} various time intervals. The residuals between the observed data and the model are shown in the lower section of the panel. Panel b: Scaled light curves in multiple energy bands are compared with the synchrotron model predictions (colored curves) and observed data (grey curves), with the dashed black line indicating the time when electron injection ceases. Panel c: The evolution of the peak energy ($E_{\rm{p}}$) over time is displayed, with blue points showing results from the CPL fits and the orange curve representing the synchrotron model fit. Panel d: Evolution of the electron energy distribution ($N_{\rm{e}}$) is shown over source-frame time, illustrating how the electron population evolves with cooling. Panel e: The time-resolved \textcolor{black}{model} energy spectra ($\nu F_\nu$) are plotted as a function of observer-frame time, showing the change in spectral shape across different time intervals.}
 \label{fig: sych}
\end{figure*}

 \begin{table}[ht!]
 \centering
\caption{The prior bounds and best-fit results of the time-evolving synchrotron model parameters used for fitting GRB 230812B.}
 \label{tab: sych}
 \begin{tabular}{ccc}
 \toprule
 \toprule
 Parameters&Prior bounds & Best fitting results\\
 \midrule
 $\log (B_{\rm{0}}/\rm{G})$&[0, 3] & $2.81^{+0.24}_{-0.13}$\\
 $\alpha_{\rm{B}}$ & [1, 2] & $1.95^{+0.07}_{-0.02}$\\
 $\log \gamma_{\rm{min}}$&[4, 7] & $5.16^{+0.01}_{-0.02}$\\
 $\log \Gamma$&[1.2, 3] & $2.79^{+0.23}_{-0.17}$\\
 $p$ & [1.5, 3.5] & $3.49^{+0.02}_{-0.01}$\\
 $t_{\rm{inj}}$/s& [-1.5, 2] & $0.86^{+0.01}_{-0.02}$\\
 $q$ & [0, 10] & $2.22^{+0.08}_{-0.05}$\\
 $\log (R_{\rm{0}}/\rm{cm})$ & [12, 16] & $14.51^{+0.42}_{-0.30}$\\
 $\log (Q_{\rm{0}}/\rm{s^{-1}})$ & [31, 156] & $48.34^{+0.50}_{-0.95}$\\
 \bottomrule
 \end{tabular}
\end{table}

\subsection{A Physical Model Fit}\label{physical model}

We observe that the overall spectral shape from the time-integrated fit, with $\alpha \sim -0.95$, along with the hard-to-soft evolution of $E_{\rm{p}}$ from the time-resolved spectral analysis, aligns with a non-thermal, synchrotron-like spectrum. The apparent violation of the synchrotron ``death-line'' \citep{Preece1998} for certain $\alpha$ values could be due to the empirical CPL model's inability to precisely capture the physical spectral curvature. Moreover, data uncertainties may still allow for $\alpha$ values below the death-line, particularly considering the forward-folding matrix response procedure employed during the fit \citep{Zhang2016}.

Above facts, combined with the single Fast-Rise Exponential-Decay (FRED) shape of the light curve, motivated us to apply a physical model fit using the synchrotron model developed by \citet{Yang2023} and \citet{Yan2024}. This model employs a unified synchrotron scenario, assuming a single electron injection event, and traces the evolution of cooling electrons in a decaying magnetic field. By tracking the time-dependent observed spectra, the model fits the time-resolved spectral data and yields a single set of physical parameters upon a successful fit.

The unified synchrotron model, expressed in units of $\rm{ph\ s^{-1}\ cm^{-2}\ keV^{-1}}$, is described as \citep{Yang2023, Yan2024}:

\begin{equation}
 \begin{split}
       F_{\nu, \rm{obs}} = \\ F_{\nu, \rm{obs}}(t, \nu, B_0, \alpha_{\rm{B}}, \gamma_{\min}, \Gamma, p, t_{\rm{inj}}, R_0, Q_0, \gamma_{\max}, z),    
 \end{split}
\end{equation}  

where the flux is calculated numerically at any given time ($t$), energy ($\nu$), and redshift ($z$) based on the parameter set $P = \{B_0, \alpha_{\rm{B}}, \gamma_{\min}, \Gamma, p, t_{\rm{inj}}, R_0, Q_0, \gamma_{\max}\}$. We applied the same fitting strategy outlined in \citet{2024ApJ...962...85Y}. During the fitting process, the maximum Lorentz factor, $\gamma_{\max}$, was fixed at $10^8$, resulting in a model characterized by nine free parameters. These parameters include $R_{\rm{0}}$, representing the initial radius of the emission region; $B_{\rm{0}}$, the magnetic field strength at $R_{\rm{0}}$; $\alpha_{\rm{B}}$, $p$, and $q$, which denote the power-law decay index of the magnetic field, the electron spectrum, and the electron injection rate, respectively; and $\Gamma$, the bulk Lorentz factor. $t_{\rm{inj}}$ and $Q_{\rm{0}}$ represent the injection time and the electron injection rate coefficient, respectively.

These parameters, constrained within the bounds listed in Table \ref{tab: sych}, were used to fit the model to the observed spectra at time $t$:

\begin{equation} \label{equ
} F_{\nu, \rm{obs}} = F_{\nu, \rm{obs}}(t, \nu, P). \end{equation}

To perform the fit, we employed the custom-developed Python tool MySpecFit \citep{Yang2022}, designed for Bayesian parameter estimation. MySpecFit is built on PyMultinest \citep{Bunchner2014}, which provides a Python interface for the Multinest nested sampling algorithm \citep{Feroz2008, Feroz2009, Feroz2019}.

After obtaining a successful fit with the time-dependent synchrotron model for GRB 230812B, indicated by statistically acceptable goodness-of-fit values (PGSTAT/d.o.f $\sim$ 1), the best-fit parameters and their 1$\sigma$ uncertainties are presented in Table \ref{tab: sych}.

Figure \ref{fig: sych} illustrates the detailed fitting results. In Figure \ref{fig: sych}a, the observational data are compared with the model predictions. The model successfully reproduces the light curves, as shown in Figure \ref{fig: sych}b, where we also calculate the predicted light curve in the 0.5–4 keV energy range of the Wide-field X-ray Telescope (WXT) onboard the Einstein Probe mission \citep{Liu2024}. Notably, the duration of GRB 230812B in the soft X-ray band exceeds that observed in the gamma-ray band, consistent with other GRB observations by EP/WXT \citep{Liu2024, Yin2024}. The model also accurately captures the hard-to-soft spectral evolution, as shown in Figures \ref{fig: sych}b and \ref{fig: sych}c. Additionally, Figures \ref{fig: sych}d and \ref{fig: sych}e illustrate the evolution of the electron distribution and the predicted $\nu F_{\nu}$ spectra.

The posterior probability distributions of these parameters are visualized in the corner plot (Figure \ref{fig: corner}). Among the fitted parameters, the estimated value of $R_{\rm{0}} \sim 10^{14.51}$ cm is consistent with theoretical expectations of the large emission radius in the framework of supernova-associated GRBs, typically originating from the collapse of a massive star.

\section{Placement of GRB 230812B}\label{placement}

\begin{figure}
 \label{fig: lum_t}
 \centering
 \includegraphics[width = 0.45\textwidth]{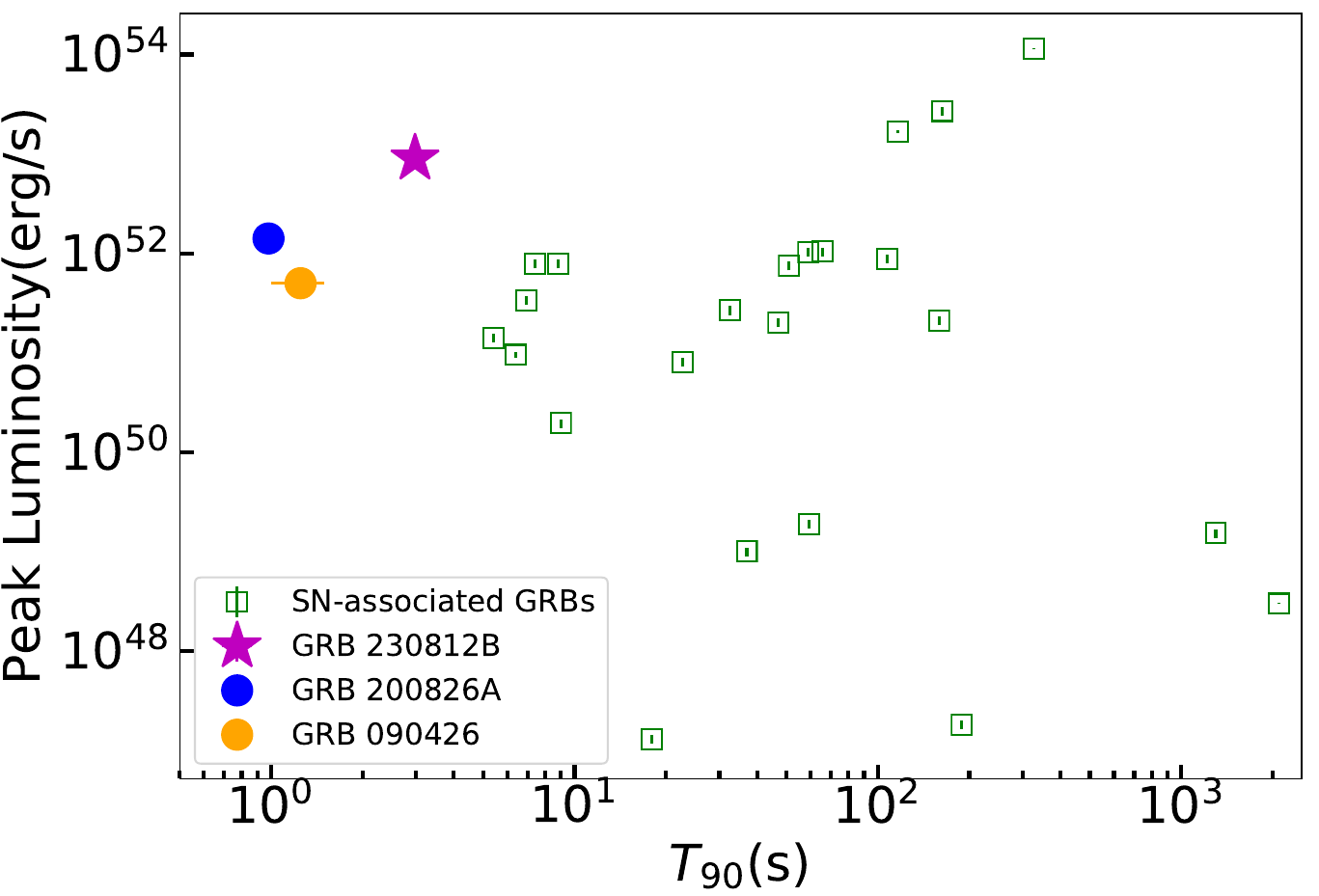}
 \caption{Peak luminosity versus $T_{\rm{90}}$ duration for GRB-SNe \citep[data from][]{Frontera2024, An2023, Li2023, Klose2018, Blustin2006, Guidorzi2005, Sazonov2004, Granot2003}. The green squares represent supernova-associated GRBs, while \textcolor{black}{GRB 230812B, GRB 200826A, and GRB090426 are indicated with the magenta, blue, and orange points, respectively.}}

\end{figure}

The detection of the associated supernova strongly indicates that GRB 230812B originates from the collapse of a massive star. In Figure \ref{fig: lum_t}, we present the peak luminosity versus $T_{\rm{90}}$ for all supernova-associated GRBs, demonstrating that GRB 230812B is a unique event, bridging the gap between those short-duration ($<2$ s) and long-duration ($>2$ s) Type-II GRBs. In this section, we further investigate its collapse origin by comparing its observed properties with those of other known Type I and Type II GRBs.

\subsection{$E_{\rm{p},z}-E_{\gamma,\rm{iso}}$ Relation}

\begin{figure}[ht!]
 \centering
 \includegraphics[width = 0.5\textwidth]{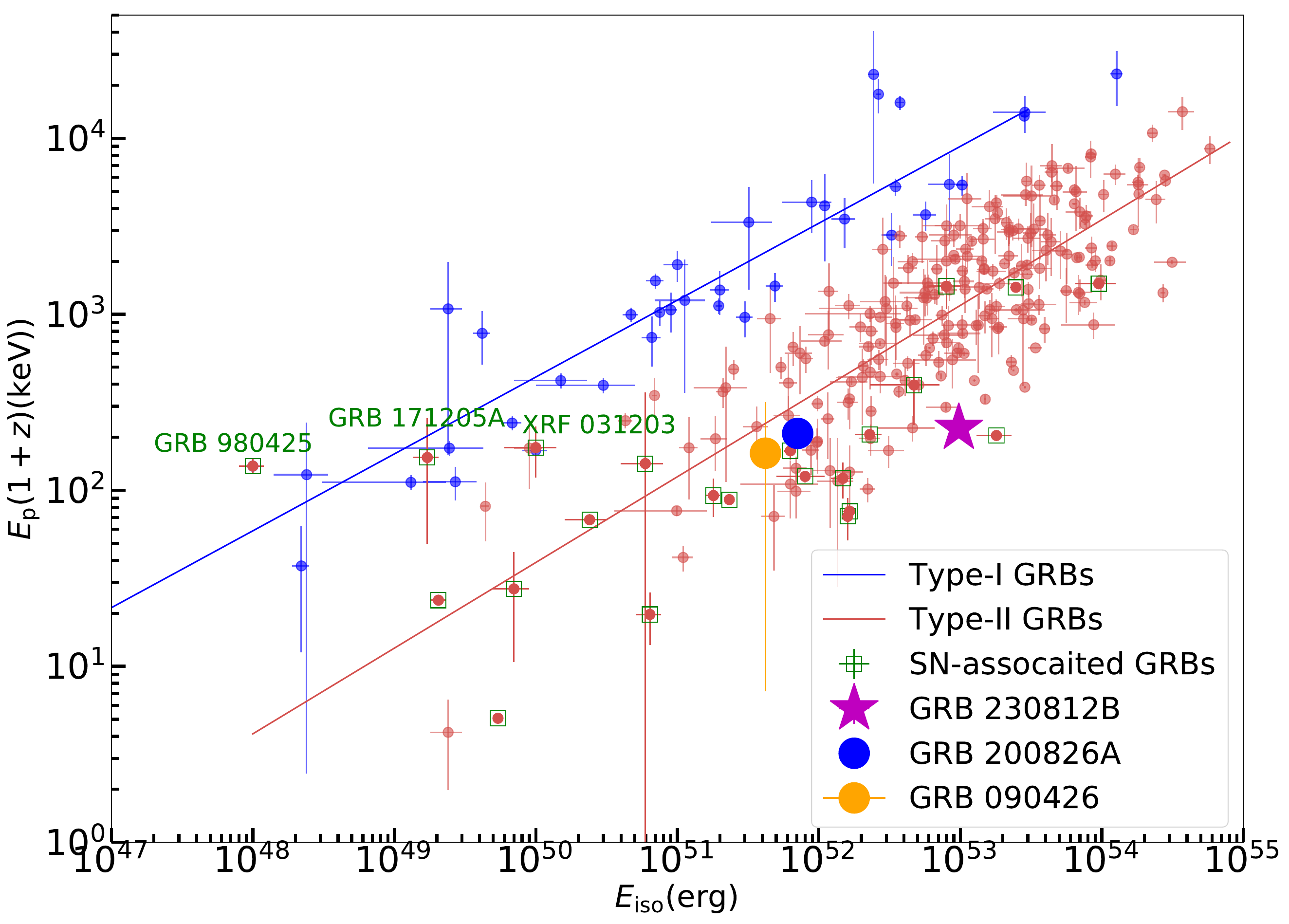}
 \caption{Placement of GRB 230812B in the Amati relation, the $E_{\rm{p},z}$–$E_{\gamma, \rm{iso}}$ correlation diagram for Type I and Type II GRBs. Type I GRBs are represented by blue circles, while Type II GRBs are marked by red circles. Supernova-associated GRBs are shown as green squares. GRB 230812B is represented by the magenta star, GRB 200826A by the blue circle, and GRB 090426 by the orange circle. The solid blue and red lines represent the Amati relation fits for Type I and Type II GRBs, respectively.}

 \label{fig: Amati}
\end{figure}

With the measured redshift of GRB 230812B, we calculated an isotropic energy release of $E_{\gamma,\rm{iso}} = 6.10^{+0.11}_{-0.54} \times 10^{53}\ \rm{erg}$. This value was derived based on the observed fluence of $8.79^{+0.08}_{-0.13} \times 10^{-4}\ \rm{erg\ cm^{-2}}$. This isotropic energy is significantly higher than typical Type I GRBs but is consistent with the energy range observed in Type II GRBs. When plotted on the $E_{\rm{p},z}-E_{\gamma,\rm{iso}}$ diagram—where $E_{p,z}$ represents the peak energy in the burst’s rest frame, calculated as $E_{\rm{p}}(1+z)$—GRB 230812B aligns with the so-called Amati Relation \citep{Amati2002}. While long and short GRBs usually trace distinct paths in this relation, GRB 230812B clearly follows the track characteristic of Type II GRBs, as shown in Figure \ref{fig: Amati}.

\subsection{Other Correlations}

In Figure \ref{fig: Corr0}, we present GRB 230812B in four different correlations commonly utilized to infer the physical classification of GRBs:

\begin{figure*}
 \centering
 \includegraphics[width = 0.77\textwidth]{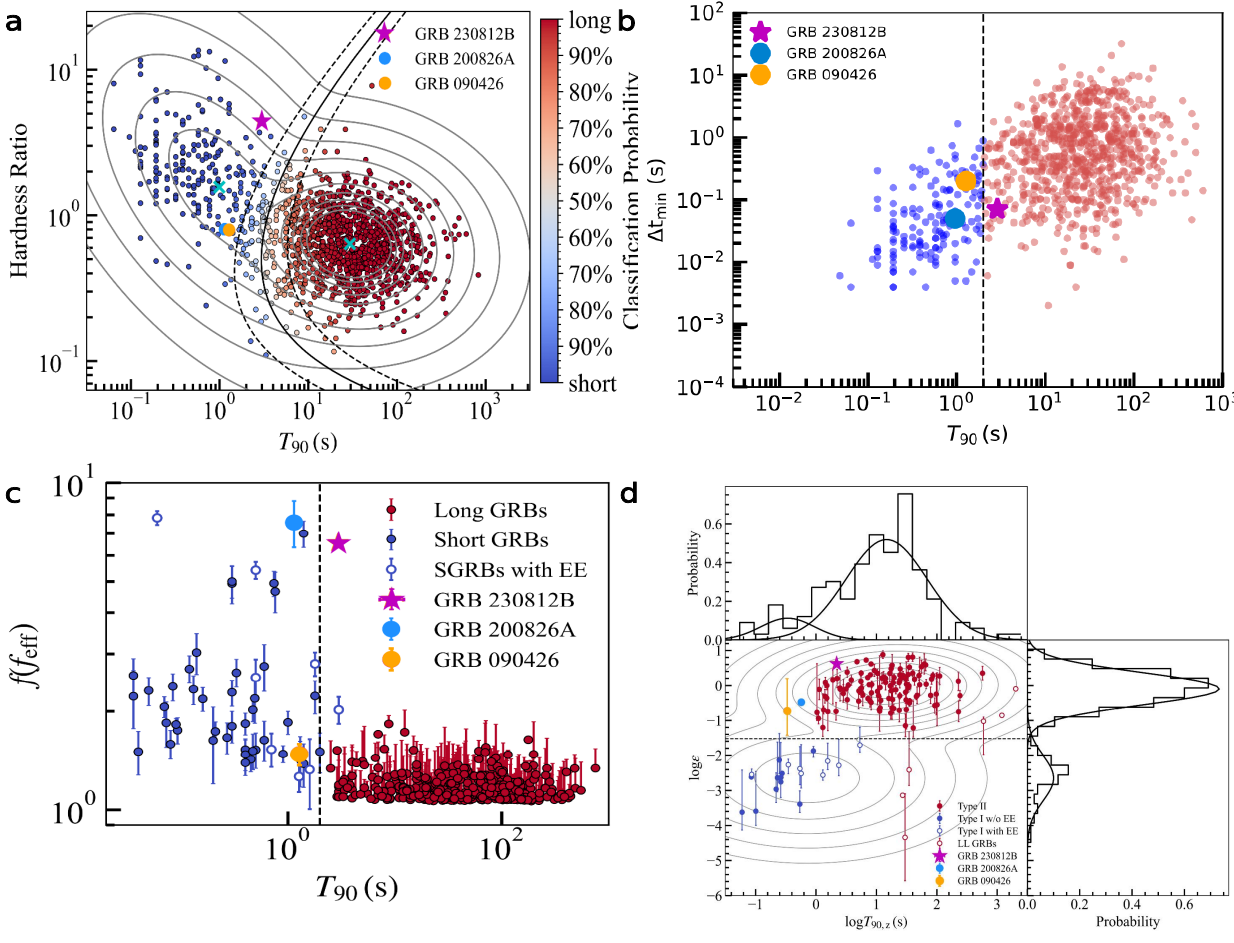}
 \caption{Placement of GRB 230812B in different correlation plots. In all panels, GRB 230812B is shown as a magenta star, GRB 200826A as a blue circle, and GRB 090426 as an orange circle. a, The $T_{\rm{90}}$–Hardness Ratio (HR) diagram. GRB 230812B is plotted against Type I GRBs (blue) and Type II GRBs (red). The black solid line at $T_{\rm{90}} = 2$ s distinguishes long GRBs (right) from short GRBs (left). The region between black dashed lines represents GRBs with ambiguous classification. b, The $\Delta t_{\rm{min}}-T_{\rm{90}}$ plot compares GRB 230812B with Type I GRBs (blue) and Type II GRBs (red).  A black dotted line marks the boundary at $T_{\rm{90}} = 2$ s. c, The $T_{\rm{90}} - f_{\rm{eff}}$ plot, where $f_{\rm{eff}}$ values for Type II GRBs are indicated by red circles, and $f$ values for Type I GRBs are indicated by blue circles. GRB 230812B stands out with its high $f_{\rm{eff}}$, supporting its classification as a short-duration Type II GRB. d, The $\epsilon-T_{\rm{90}}$ plot, illustrating the parameter $\epsilon$ in relation to $T_{\rm{90}}$. The Type I and Type II GRB candidates are shown by blue and red circles, respectively. Gaussian fits for the $\epsilon$ and $T_{\rm{90}}$ distributions are overlaid in black.}
 \label{fig: Corr0}
\end{figure*}

\begin{itemize}
   \item \textbf{HR-$T_{\rm{90}}$ Correlation}: The hardness ratio (HR) is defined as $\rm{HR} = F_{50-300}/F_{10-50}$, where $F_{50-300}$ and $F_{10-50}$ represent the fluxes in the 50–300 keV and 10–50 keV energy ranges, respectively. Statistically, Type II GRBs tend to exhibit lower HR values, reflecting a softer energy profile compared to Type I GRBs. In the HR-$T_{\rm{90}}$ plot (Figure \ref{fig: Corr0}a), GRB 230812B, with an HR of approximately 4.44, stands out as an outlier alongside GRBs 200826A and 090426, primarily due to their short durations, which set them apart from the typical Type II GRBs.

  \item \textbf{Minimum Variability Timescale ($\Delta t_{\rm{min}}$)-$T_{\rm{90}}$ Correlation}: The minimum variability timescale, $\Delta t_{\rm{min}}$, represents the shortest interval over which a GRB exhibits significant flux variations. This parameter provides valuable insights into the size of the radiation zone, the Lorentz factor, and the emission mechanism \citep{Schmidt1987}. Using the Bayesian block method \citep{Scargle2013}, we segmented GRB 230812B’s light curve into distinct emission episodes and determined a minimum variability timescale of $71^{+16}_{-2}$ ms (Figure \ref{fig: Corr0}b). This timescale places GRB 230812B in an intermediate range between the two GRB types \citep{Camisasca2023}, further supporting its role as a bridge between typical long-duration Type II GRBs and peculiar short-duration Type II GRBs.

  \item \textbf{$f$-Parameter-$T_{\rm{90}}$ Correlation}: The $f$ parameter, defined as $f \equiv \frac{F_{\rm{p}}}{F_{\rm{b}}}$, compares the peak flux ($F_{\rm{p}}$) to the background flux ($F_{\rm{b}}$) for short GRBs. For long GRBs, this is re-scaled as $f_{\rm{eff}} = \frac{F_{\rm{p}}'}{F_{\rm{b}}'}$ to account for the tip-of-the-iceberg effect, which can truncate a long GRB's observed duration \citep{Lu2014}. This correlation distinguishes between intrinsically short GRBs and long GRBs with truncated durations. GRB 230812B shows an $f_{\rm{eff}}$ value of $6.54 \pm 0.10$, well above the critical threshold of 1.5, confirming that its short duration is intrinsic and not due to the tip-of-the-iceberg effect (Figure \ref{fig: Corr0}c), similar to the case of GRB 200826A.

    \item \textbf{$\epsilon$-$T_{\rm{90}}$ Correlation}: The $\epsilon$ parameter, calculated as $\epsilon = E_{\gamma, \rm{iso, 52}} / E_{\rm{p,}z\rm{,2}}^{5/3}$, where $E_{\gamma,\rm{iso,52}}$ is the isotropic gamma-ray energy in units of $10^{52}$ erg and $E_{\rm{p,z,2}}$ is the rest-frame peak energy normalized to 100 keV, serves as a classification tool between Type I and Type II GRBs \citep{Zhang2006}. With an $\epsilon$ value of 4.21, GRB 230812B clearly falls within the Type II GRB regime (Figure \ref{fig: Corr0}d), well above the 0.03 threshold that separates the two classes \citep{Lu2010}.

\end{itemize}

The analysis of GRB 230812B across multiple correlations above confirms its consistency with the properties of a Type II GRB, aligning with its supernova association. However, its unusually short duration, much like GRB 200826A, sets it apart as an outlier within the typical Type II sample when $T_{\rm{90}}$ is considered. This short duration appears to be an intrinsic feature of the burst rather than an observational artifact, suggesting it represents a fundamental characteristic that requires further exploration.

\subsection{GRB 230812B as a Twin Case of GRB 200826A}

Having established that both GRB 230812B and GRB 200826A are simiar outliers within the Type II GRB population due to their unusually short durations, we now focus on the striking similarities between these two events. Like GRB 230812B, GRB 200826A is also associated with a supernova and exhibits a instrinsic short duration, challenging the conventional classification of supernova-associated GRBs as long-duration events. This section explores these parallels, suggesting that GRB 230812B and GRB 200826A may represent a continuous sub-class of supernova-associated GRBs with exceptionally short durations.

\begin{figure}[ht!]
 \centering
 \includegraphics[width = 0.5\textwidth]{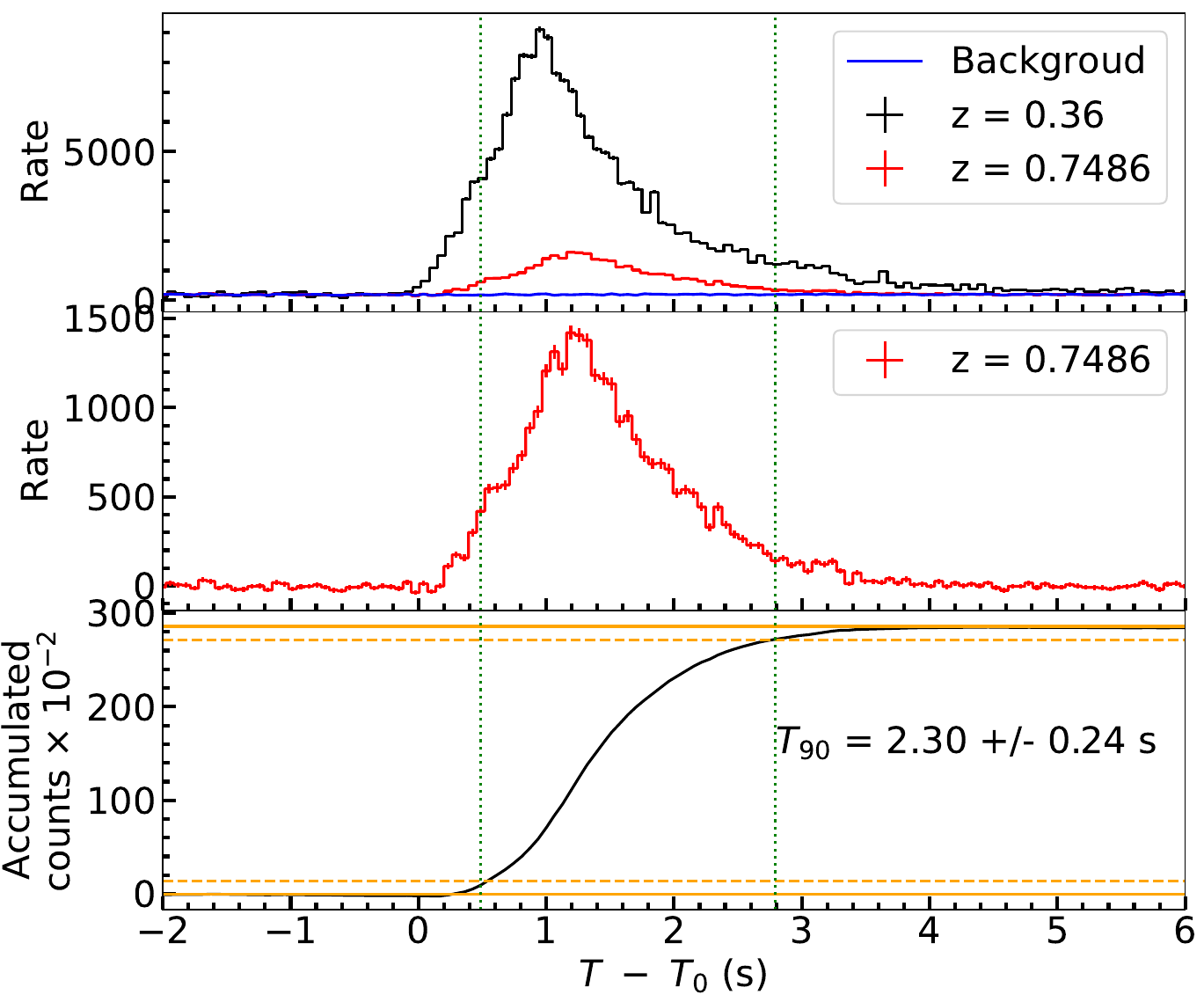}
 \caption{Simulated effect of increased redshift on the observed duration of GRB 230812B. \textit{Top panel}: The black curve represents the original light curve of GRB 230812B at its actual redshift ($z=0.36$). The red curve shows the simulated light curve if GRB 230812B were observed at a higher redshift ($z=0.7486$), the same as GRB 200826A. The blue line indicates the background level. \textit{Middle panel}: A zoomed-in view of the simulated light curve at $z=0.7486$ (red curve). \textit{Bottom panel}: The $T_{\rm{90}}$ calculation for the simulated GRB 230812B at $z=0.7486$, showing that the observed duration decreases to approximately 2.3 seconds. This reduction in $T_{\rm{90}}$ would categorize GRB 230812B more like a short-duration GRB if it were at the same redshift as GRB 200826A, demonstrating the impact of redshift on burst classification.}

 \label{fig: mr}
\end{figure}

To  demonstrate the similarity between these two bursts, we recalculated the $T_{\rm 90}$ of GRB 230812B by placing it at the same redshift as GRB 200826A \citep[$z=0.7486$, ][]{Rossi2022}. The scaled flux and time relations indicate that much of GRB 230812B's signal would be submerged in background noise at this greater distance, thereby reducing its observed duration. Using the relation ${\rm Flux}(z_1) = {\rm Flux}(z_0)\times \left(\frac{D_{L,z_0}}{D_{L,z_1}}\right)^2$, where $z_0 = 0.36$ for GRB 230812B and $z_1 = 0.7486$ for GRB 200826A, we infer that if both bursts were observed at the same distance ($z = 0.7486$), the $T_{\rm{90}}$ of GRB 230812B would decrease to $\sim$ 2.3 seconds, thus categorizing it more like a short-duration burst (Figure \ref{fig: mr}).

This analysis strongly suggests that GRB 230812B is a ``twin case'' of GRB 200826A. Both bursts challenge the conventional understanding that supernova-associated GRBs typically have long durations. Together, they reveal a rare sub-class of Type II GRBs with unusually short durations. GRB 230812B, following the precedent set by GRB 200826A, stands as the second known example of this phenomenon, emphasizing the need for further investigation into the mechanisms responsible for these short-duration Type II GRBs.

\section{Explanation of the Short Duration}\label{explanation}

Explaining the observed short durations, on the order of a few seconds, in GRBs 230812B presents a significant challenge within the collapsar framework. According to this model, the burst duration should correspond to the minimum timescale for accretion onto the central engine, which is governed by the free-fall timescale of the progenitor star. This timescale is expressed as:

\begin{equation}
    t_{\rm{ff}} \approx \left( \frac{3\pi}{32G\bar{\rho}} \right)^{1/2} \approx 210 \, \text{s} \left( \frac{\bar{\rho}}{100 \, \text{g cm}^{-3}} \right)^{-1/2}, 
\end{equation}
where $\bar{\rho}$ is the mean density of the accreted material \citep{Zhang2018}.

For a typical massive star with a density of $\bar{\rho} \approx 100 \, \rm{g\ cm^{-3}}$, the expected GRB duration should be much longer than a few seconds. Additionally, the high observed $f$ value makes the tip-of-the-iceberg interpretation \citep{Levesque2009, Lu2014}, which suggests the GRB could be intrinsically longer but partially hidden by detector sensitivity, infeasible for explaining the short duration of GRB 230812B.

To address this challenge, \citet{Zhang2021} proposed two possible scenarios regarding the jet behavior in such GRBs:

\begin{itemize}
    \item \textbf{Baryon-Loaded Jet Scenario}: In this case, the jet may carry a significant amount of baryonic matter, making it only mildly relativistic or even non-relativistic \citep{Metzger2011}. As a result, the weaker, baryon-loaded jet would emit gamma rays for a shorter duration. However, such jets are not expected to produce the bright gamma-ray emission observed in GRB 230812B, making this scenario less consistent with the data.
    \item \textbf{Jet Penetration Scenario}: In this scenario, the GRB duration is affected by the time it takes for the jet to penetrate the stellar envelope. The total central engine activity timescale, denoted as $\Delta t_{\rm{eng}}$, could be longer, while the observed GRB duration, $\Delta t_{\rm{GRB}}$, would correspond to the remaining time after the jet breakout, i.e., $\Delta t_{\rm{GRB}} = \Delta t_{\rm{eng}} - \Delta t_{\rm{jet}}$. In this model, the jet spends most of its time breaking out of the stellar envelope, and the short observed GRB duration is the remainder.
\end{itemize}

In the second scenario, the total central engine timescale, $\Delta t_{\rm{eng}}$, would be consistent with the typical duration of a massive star collapse, generally exceeding 10 seconds. In the cases of GRB 230812B and GRB 200826A, the jet breakout timescale, $\Delta t_{\rm{jet}}$, is only slightly shorter than $\Delta t_{\rm{eng}}$. Specifically, for GRB 200826A, this difference is about 1 second, resulting in a GRB duration of approximately 1 second, while for GRB 230812B, the difference is around 3 seconds, producing a GRB duration of approximately 3 seconds. This interpretation requires a coincidence in the alignment between $\Delta t_{\rm{eng}}$ and $\Delta t_{\rm{jet}}$, a possibility noted in previous research \citep{Ahumada2020}.

The case of GRB 230812B, along with GRB 200826A, suggests that short-duration supernova-associated GRBs are not isolated incidents. The consistency between these two GRBs reinforces the feasibility of the jet penetration model. Further exploring the variation between $\Delta t_{\rm{eng}}$ and $\Delta t_{\rm{jet}}$ could clarify the relationship between these timescales. For GRBs whose durations are close to $\Delta t_{\rm{eng}}$, the jet breakout time may be relatively insignificant, leading to longer bursts. Conversely, shorter GRB durations like those observed in GRB 230812B and GRB 200826A occur when the difference between these two timescales is minimal.

This model offers a compelling explanation for GRBs with $T_{\rm{90}}$ near the boundary between Type I and Type II classifications. It suggests that supernova-associated GRBs with durations under 10 seconds could naturally arise from this alignment of engine and jet timescales, as illustrated in Figure \ref{fig: t90_hist}.

\section{Summary}\label{summary}

In this study, we analyzed the short-duration GRB 230812B, detected by the GRID mission. Despite its relatively short duration of approximately 3 seconds, GRB 230812B is associated with a supernova, challenging the conventional view that short GRBs arise exclusively from compact object mergers. Temporal and spectral analyses show that GRB 230812B exhibits typical Type II GRB characteristics, including significant spectral lags and a hard-to-soft spectral evolution, which align with the collapse of a massive star.

The spectral analysis supports a synchrotron emission model with an emission radius of around $10^{14.5}$ cm, consistent with the collapse of a massive star. We explored possible explanations for the burst's short duration, focusing on the interplay between the central engine's activity and the jet's penetration time through the stellar envelope. This model suggests that the short duration may result from a small difference between the total engine timescale and the jet breakout time.

Comparisons with GRB 200826A suggest that GRB 230812B may belong to a distinct subclass of supernova-associated GRBs characterized by unusually short durations. Both bursts exhibit durations significantly shorter than the typical timescales for massive star collapses, yet their supernova associations point to a progenitor involving stellar collapse rather than compact object mergers. This challenges the traditional classification of GRBs into long and short categories based on duration alone and highlights the need to consider the central engine’s activity and jet dynamics in interpreting GRB durations. These findings provide new insights into the diversity of Type II GRBs and suggest that short-duration GRBs may not be exclusive to compact object mergers, but could also arise from massive star collapses under certain conditions.

\begin{acknowledgments}
We acknowledge the support from the National Key Research and Development Programs of China (2022YFF0711404, 2022SKA0130102), the National SKA Program of China (2022SKA0130100), the National Natural Science Foundation of China (Grant Nos. \textcolor{black}{12025301}, 11833003, U2038105, U1831135, 12121003), the science research grants from the China Manned Space Project with NO.CMS-CSST-2021-B11, the Fundamental Research Funds for the Central Universities, the Program for Innovative Talents and Entrepreneurs in Jiangsu, and \textcolor{black}{the Strategic Priority Research Program of the Chinese Academy of Sciences}. This work was performed on an HPC server equipped with two Intel Xeon Gold 6248 modules at Nanjing University. We acknowledge IT support from the computer lab of the School of Astronomy and Space Science at Nanjing University.
\end{acknowledgments}

\bibliography{ms}
\bibliographystyle{aasjournal}

\newpage 

\appendix
\counterwithin{figure}{section}
\counterwithin{table}{section}
\counterwithin{equation}{section}
\section{The posterior probability distributions of  best-fit parameters of the synchrotron model}
\counterwithin{figure}{section}
\counterwithin{table}{section}
\counterwithin{equation}{section}

\begin{figure}[h!]
 \centering
 \includegraphics[width = 0.86\textwidth]{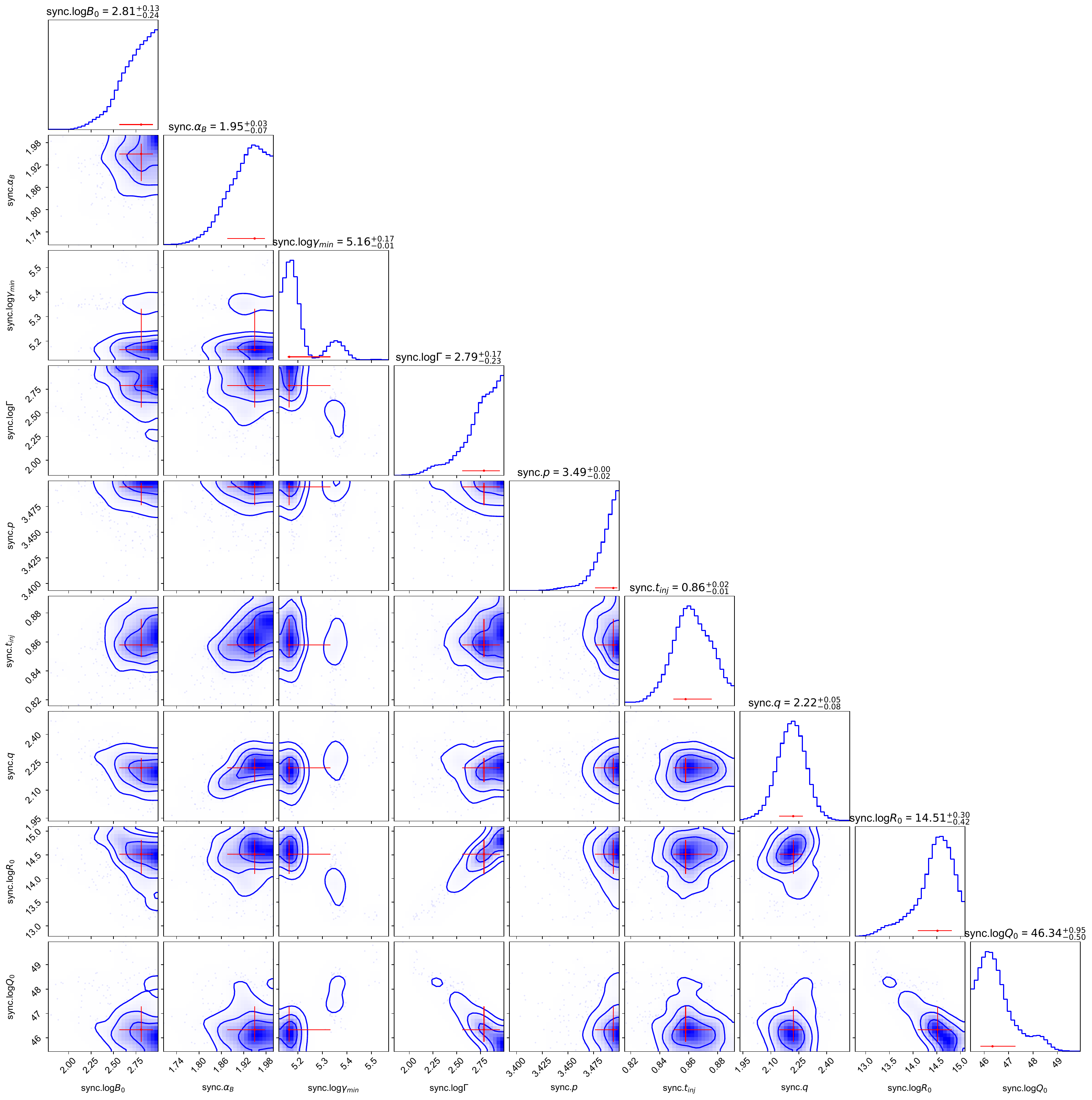}
 \caption{Corner plot of the posterior probability distributions for the parameters from the synchrotron model fitted to the time-evolved spectra. The red error bars denote 1$\sigma$ uncertainties.}
 \label{fig: corner}
 \end{figure}

% \begin{figure}[h!]
% \plotone{FigureC1.pdf}
% \caption{Same as Figure \ref{fig:120326A}, but for GRB 131011A.
% \label{fig:131011A}}
% \end{figure}

\end{document}